\documentstyle[prd,aps,epsf,epsfig]{revtex} 
\begin{document}
\draft
\twocolumn[\hsize\textwidth\columnwidth\hsize\csname@twocolumnfalse\endcsname
%
\title{Flatness of the setting Sun}
\vspace{0.5cm}
\author{Z. N\'eda and S. Volk\'an}
\address{Babe\c{s}-Bolyai University, Dept. of Physics, RO-3400, Cluj, Romania \\
E-mail: zneda@phys.ubbcluj.ro}

\maketitle
\centerline{\small (Last revised \today)}

\begin{abstract}
Atmospheric refraction is responsible for the bending of
light-rays in the atmosphere. It is a result of the continuous
decrease in the refractive index of the air as a function of
altitude. A well-known consequence of this phenomenon is the
apparently elliptic shape of the setting or rising Sun (or
full-Moon). In the present paper we systematically investigate
this phenomenon in a standard atmosphere. Theoretical and
numerical calculations are compared with experimental data. The
asymmetric rim of the Sun is computed as a function of its
inclination angle, observational height and meteorological
conditions characterized by pressure, temperature and lapse-rate.
We reveal and illustrate some extreme and highly unusual
situations.
\end{abstract}

\vspace{2pc}
]

\vspace{1cm}


\section{Atmospheric refractions}

Atmospheric refraction is between the first scientifically
approached phenomenon. This effect is responsible for the apparent
scintillation of stars, mirages, the puzzling and spectacular
green flash, difference between apparent and real position of
stars or the asymmetric rim of the Sun near the horizon. The
refractive index of dry air is very close to $1$. Its small
dependence on temperature and pressure leads to a refractive index
gradient in the atmosphere. Although this gradient is very small,
due to the large distances travelled by the light-rays in the
atmosphere, it can result observable bending or dispersion.

Atmospheric refraction is usually divided into three categories:
astronomical, terrestrial and geodesic. Terrestrial refraction
appears when both the object and observer are within the Earth's
atmosphere. This refraction is responsible for ordinary mirages,
and has been extensively studied \cite{mirages}. The first one
reporting and accounting for mirages was Aristotle in
Meteorologica \cite{aristotle}. Correctly, he concluded that dense
air layers can act as mirrors, and considered this effect
responsible for mirages. (A complete historic and bibliographic
study of mirages can be found on the splendid web-page of
A.T.Young \cite{young}.) Geodesic refraction is a special case of
terrestrial refraction where both the object and observer are at
low altitudes. A well-known example for this is surveying. We
speak about astronomical refraction when a terrestrial observer
detects ray-bending effects for the light-rays coming from objects
outside the Earth's atmosphere. This refraction is responsible for
the difference between the real and apparent position of stars
near the horizon, the green flash or the asymmetric rim of the
setting (or rising) Sun. In the present paper we discuss in detail
the flatness of the Sun or full-Moon near the horizon.
Particularly, we are interested in the extreme values this
flatness can take for standard atmospheric profiles, and the
dependence of the flatness on the altitude of the observer and
meteorological conditions (temperature and pressure). A similar
study, considering fixed atmospheric conditions was recently
published by Thomas and Joseph \cite{hopkins}. Here we plan a more
complete analysis, considering three different theoretical
methods, experiments, computer simulations and illustration of the
results with a collection of pictures and movies.

The spectacular nature of sunset or sunrise makes the phenomenon
attractive to students. After our experience it can effectively be
used to exemplify refraction phenomena and exercising the
principles of geometrical optics. Our paper is structured as
follows. First, we introduce the atmosphere model, and the
refractive index profile is given. Then, we present three
different methods for computing the ray-path in this atmosphere.
Results computed for different meteorological conditions and
observation altitude will close the theoretical part.
Experimentally, we analyze and compare with theory the measured
flatness for some sunsets photographed or video-filmed by us.
Finally, we discuss some extreme and unusual conditions which are
exemplified by pictures and movies on a web-page supporting this
study. From this web-page one can also freely download a
computer-program written by us, which visualizes the sunset for
arbitrary meteorological conditions and observation altitude.

We emphasize here, that our study is restricted to standard
atmospheric conditions with smooth temperature and pressure
profile. Non-standard, but often encountered atmospheric profiles
leads to non-standard distortions of the solar rim, and are not
considered within this study.

\section{The optical atmosphere model (refractive index as a function of altitude)}

The atmosphere of the Earth is composed mainly of $N_2$ ($79\%$)
and $O_2$ ($20\%$), and extends up to a few hundreds kilometer
height. The refractive index of air is very close to $1$, and
depends slightly on its pressure and temperature following Edlen's
semi-empirical law \cite{edlen}, valid for dry air:
\begin{equation}
n=1+10^{-6}(776.2+4.36 \times 10^{-8} \nu^2)\frac{P}{T}.
\label{edlen}
\end{equation}
In the above formula $\nu$ is the wave-number of the light in
$cm^{-1}$, $P$ is the pressure in $kPa$ and $T$ the temperature in
$K$. Since the pressure and temperature varies within the
atmosphere, we get a refractive index gradient which is
responsible for atmospheric refractions. Although the variations
in $n$ are quite small, the large distances travelled by
light-rays in the atmosphere makes refraction effects observable
and sometimes important.

\begin{figure}[htb]
\epsfig{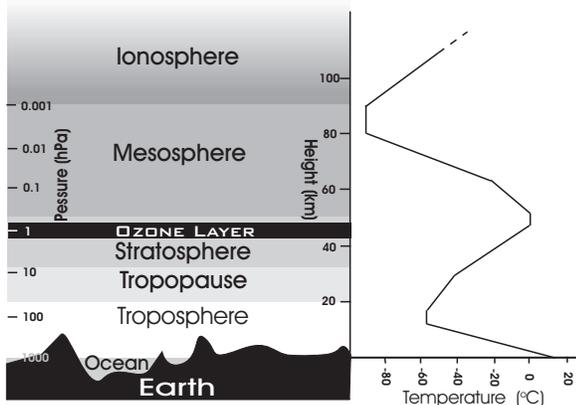} \caption{Distinct
layers of the atmosphere, and the standard temperature and
pressure profile.} \label{fig1}
\end{figure}

As a function of altitude several layers with different physical
properties are distinguishable. The lowest layer extending from
sea-level to approximative $z_t=14km$ height is called the
troposphere. This is the region where the wheatear takes place,
i.e. the region of rising and falling packets of air. In this
layer the air pressure drops drastically, at the top of the
troposphere being only $10\%$ of the value measured at sea-level.
In the troposphere the temperature decreases almost linearly with
the altitude (Fig.~1). A thin buffer zone between the troposphere
and the next layer (the stratosphere) is called the tropopause.
Within the tropopause the temperature is in good approximation
constant. The stratosphere extends from the altitude of
approximately $18km$ up to $z_s=50km$ (Fig.~1). Within the
stratosphere the air flow is mostly horizontal. In the upper part
of the stratosphere we have the ozone layer. Inside the
stratosphere the pressure decreases further with the altitude, but
surprisingly the temperature increases with height. Above the
stratosphere we find the mesosphere and the ionosphere. In these
regions the air is very rare and the temperature profile is shown
in Fig.~1.

From the viewpoint of atmospheric refractions only the first two layers, the troposphere
and the stratosphere are important. In the upper layers of the atmosphere the air is
so rarefied, that the refractive index can be considered $1$ within a good approximation.

An accepted and widely used model for the atmosphere is the U.S.
Standard Atmosphere, established in 1953 and re-actualized in 1976
\cite{USstandard}. The U.S. Standard Atmosphere consists of single
profiles, representing the idealized steady-state atmosphere for
moderate solar activity. The listed parameters include
temperature, pressure, density, gravitational acceleration, mean
particle speed, mean collision frequency, mean free path, etc. as
a function of altitude. In our study we consider an atmosphere
model with a spherical symmetry, all relevant physical quantities
(temperature and pressure) varying only as a function of altitude.
We then calculate the refraction index of air as a function of altitude as follows: \\
(i) The temperature profile in the troposphere is linearly
decreasing with a $\lambda=6.5K/km$ lapse-rate, as suggested by
the U.S. Standard Atmosphere. Following again the U.S. Standard
Atmosphere we consider the temperature constant within the
tropopause. Although the temperature increases as a function of
altitude in the stratosphere, due to the rarefied air (small
pressure) the refractive index is close to $1$, and for the sake
of simplicity we assume the temperature as constant in this
region, too. (We checked that this approximation is fully
justified.) Above $z_s=50km$ height we assume that the refractive
index is $1$, and do not calculate it anymore by Edlen's formula.
Up to the top of the stratosphere the temperature profile is
therefore presumed as
\begin{eqnarray}
T(z)= T_0-\lambda z  \: \: for \: \: z<z_t
\label{Tprofile1} \\
T(z)=T_0-\lambda z_t  \: \: for \: \: z_t\le z \le z_s,
\label{Tprofile2}
\end{eqnarray}
where $z$ is the altitude from sea-level, and $T_0$ the temperature at sea-level. \\
(ii) For the pressure profile we use a barometric formula in which we take into
account the variation of temperature with altitude. Considering a vertical slice of air with
thickness $dz$, the variation of pressure within this slice is due to the hydrostatic pressure
\begin{equation}
dP=-\rho(z) \: g(z) \: dz,
\label{dp}
\end{equation}
where $g(z)$ is the gravitational acceleration and $\rho(z)$ the density of air at height $z$:
\begin{equation}
\rho(z)=\frac{M}{V}=\frac{Nm}{V}=\frac{P(z)m}{kT(z)}.
\label{density}
\end{equation}
(We denoted by $m$ the mass of one molecule, $N$ the number of
molecules in volume $V$, $T(z)$ the temperature at height $z$ and
$k$ the Boltzmann constant.) Since we focus on the troposphere and
the stratosphere only, the $z$ height is small in comparison with
the radius of the Earth ($R_E\approx 6378km$), thus the $g$
gravitational acceleration can be considered constant. We can
write thus
\begin{equation}
dP=-\frac{P(z)mg dz}{kT(z)},
\end{equation}
which is a separable differential equation for $P(z)$. Integrating this between a height
$z_0$ where the pressure is $P_0$ and an arbitrary height $z$, we get the desired
barometric formula:
\begin{equation}
P(z)=P_0 exp\left (-\frac{mg}{k} \int_{z_0}^z \frac{dz'}{T(z')}\right ).
\end{equation}
Considering $z_0=0$ (sea-level) and using the temperature profile given by equation
(\ref{Tprofile1}) and (\ref{Tprofile2}) a simple calculus leads us to
\begin{equation}
P(z)=P_0 \left [1-\frac{\lambda z}{T_0} \right ]^{mg/k\lambda}
\label{Pprofile1}
\end{equation}
for $z<z_t$, and
\begin{equation}
P(z)= P_0 \left [1-\frac{\lambda z_t}{T_0} \right ]^{mg/k\lambda}
exp\left (-\frac{mg(z-z_t)}{k(T_0-\lambda z_t)} \right ),
\label{Pprofile2}
\end{equation}
for $z_t\le z \le z_s$.
Plugging this and (\ref{Tprofile1})-(\ref{Tprofile2}) in Edlen's formula
(\ref{edlen}), we get for the troposphere and stratosphere the variation
of the refractive index as a function of altitude. As stated before, for altitudes higher
than $z_s$ we can merely use $n=1$. The refractive index profile as a function of altitude
calculated in this manner for the parameters of the U.S. Standard Atmosphere is
plotted in Fig.~2.

\begin{figure}[htb]
\epsfig{figure=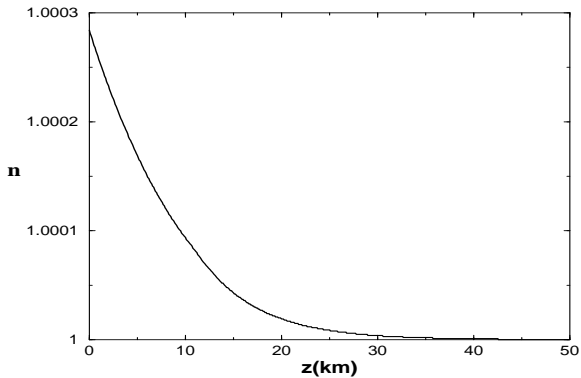,width=3.0in,height=2.0in,angle=-0}
\caption{The used refractive index profile for $T_0=10^0C$, $P_0=1atm$,
$\lambda=6.5K/km$ and $\nu=20000cm^{-1}$.}
\label{fig2}
\end{figure}

\section{Trajectory of a light-ray in the atmosphere}

We present now three different methods for computing the path of a light-ray in our
optical atmosphere model. We proved that the results given by these
methods are the same, justifying our forthcoming theoretical considerations. None of these
methods is purely analytical, they all make use of numerical calculations
to derive the light-ray trajectory. In the following we briefly describe these methods
and sketch how one can compute the deviation angle due to atmospheric refraction
for light-rays coming from distant sources. In order to avoid the phenomenon of dispersion let us
first consider monochromatic light.

\subsection{The Integral method}

This method closely follows the one discussed by Smart \cite{smart}. To illustrate the
method we use the geometry from Fig.~3.

\begin{figure}[htb]
\epsfig{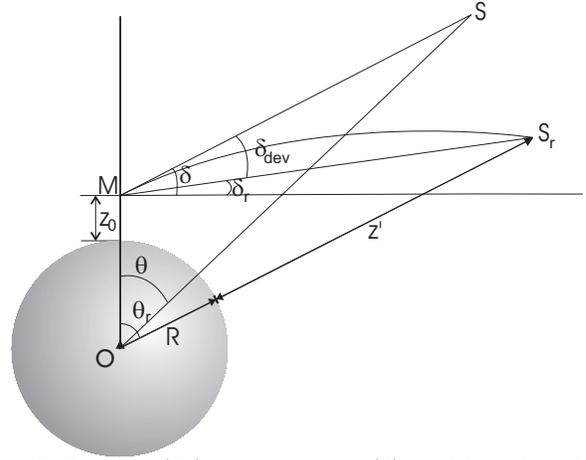}
\caption{Real ($S_r$) and apparent (S) position of a distant source $S$, as observed from
the point $M$ within the Earth's inner atmosphere.
}
\label{fig3}
\end{figure}

An observer in $M$ (at altitude $z_0$) detects the light-source placed in $S_r$ (altitude
$z'$). Let the arc $S_rM$ be the presumed path of the light-ray between the $S_r$ source and
the observer in $M$. The observer detects the light ray in the $MS$ direction, which is
tangent to the ray path in $M$. The source will be positioned by the observer in $S$
($SO$ and $S_rO$ have radial directions, thus the light-rays in these directions would not bend
within our optical atmosphere model). We denote by $\delta$ the apparent inclination angle,
characterizing the direction of the $S$ image. Let $\delta_{dev}$ be the
deviation angle of $MS$ relative to the real $MS_r$ direction of the source.
The meaning of the $\theta$ and $\theta_r$ angles are obvious from the figure. We are now interested to
compute $\delta_{dev}$ as a function of $\delta$. Presuming that $z'>>z_0$ (i.e. the source
is very far from the Earth), we can approximate $MS$, $MS_r$ and $OS$ by $z'$. Some elementary
geometry will convince us that the following approximations are justified:
\begin{eqnarray}
\theta \approx \frac{\pi}{2}-\delta-arcsin\left [\frac{z_0+R_E}{z'+R_E}\cos{\delta} \right ]
\label{approx1}\\
\delta_{dev} \approx \theta_r-\theta +\frac{R_E+z_0}{R_E+z'} (\sin{\theta_r}-\sin{\theta})
\label{approx2}
\end{eqnarray}
In order to get the desired $\delta_{dev}(\delta)$ dependence we need $\theta_r$ as a
function of $\delta$.

Let us follow now a light-ray approaching the Earth and let us
consider the atmosphere stratified in infinitesimally thin layers
of thickness $\Delta r$, with slightly different refractive
indices (Fig.~4).

\begin{figure}[htb]
\epsfig{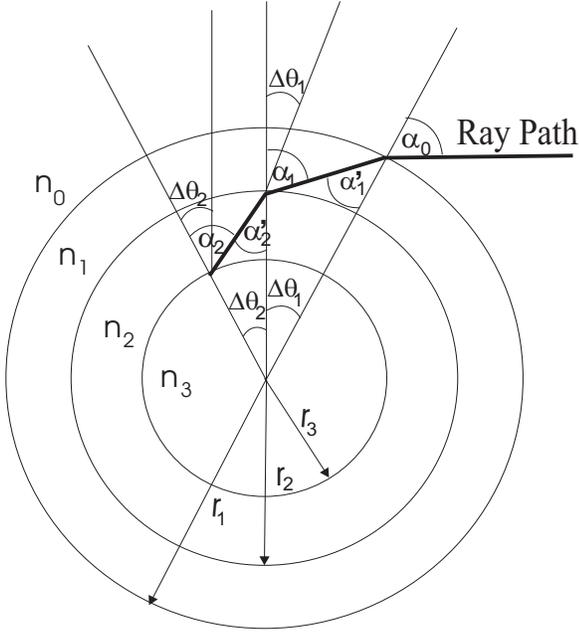}
\caption{Trajectory of a light-ray in a layered optical atmosphere model.}
\label{fig4}
\end{figure}

A useful relation between the initial incident angle $\alpha_0$
(in layer with refractive index $n_0$) and a later incident angle
$\alpha_k$ (for a layer with refractive index $n_k$) can be
derived. Using Snell's law and the notations from Fig.~4, we can
write:
\begin{eqnarray}
\frac{\sin{\alpha_0}}{\sin{\alpha_1'}}=\frac{n_1}{n_0}
\label{l1} \\
\Delta \theta_1=\alpha_1 - \alpha_1'.
\label{l2}
\end{eqnarray}
This leads to:
\begin{equation}
\sin{\alpha_1}=\frac{n_0}{n_1} \sin{\alpha_0}\left (\cos{\Delta \theta_1}+\sin{\Delta \theta_1}
\cot{\alpha_1'} \right ).
\label{l3}
\end{equation}
Similarly
\begin{equation}
\sin{\alpha_2}=\frac{n_1}{n_2} \sin{\alpha_1} (\cos{\Delta \theta_2}+\sin{\Delta \theta_2}
\cot{\alpha_2'}),
\label{l4}
\end{equation}
and combining with (\ref{l3}) it leads to:
\begin{eqnarray}
\nonumber
\sin{\alpha_2}=\frac{n_0}{n_2} \sin{\alpha_0} (\cos{\Delta \theta_1}+\sin{\Delta \theta_1}
\cot{\alpha_1'}) \times\\
(\cos{\Delta \theta_2}+\sin{\Delta \theta_2} \cot{\alpha_2'}).
\label{l5}
\end{eqnarray}
Generalizing the above equation we get:
\begin{eqnarray}
\nonumber
\sin{\alpha_k}=\frac{n_0}{n_k} \sin{\alpha_0} (\cos{\Delta \theta_1}+\sin{\Delta \theta_1}
\cot{\alpha_1'}) \times ...\times\\
(\cos{\Delta \theta_k}+\sin{\Delta \theta_k} \cot{\alpha_k'}).
\label{l6}
\end{eqnarray}
Since the $\Delta \theta_i$ angles are small, the following approximations are justified:
\begin{eqnarray}
\sin{\Delta \theta_i} \approx \Delta \theta_i,
\label{l7} \\
\cos{\Delta \theta_i} \approx  1,
\label{l8} \\
\cot{\alpha_i'} \approx \frac{\Delta r}{r_i \: \Delta \theta_i}.
\label{l9}
\end{eqnarray}
Using these approximations and the fact that $\Delta r << r_i$ we get:
\begin{eqnarray}
\nonumber \sin{\alpha_k}=\frac{n_0}{n_k} \sin{\alpha_0} \prod_{i=1}^k (1+\Delta \theta_i
\frac{\Delta r}{r_i \Delta \theta_i}) \approx   \\
\nonumber \frac{n_0}{n_k} \sin{\alpha_0} (1+\sum_{i=1}^k \frac{\Delta r}{r_i}) \approx
\frac{n_0}{n_k} \sin{\alpha_0} (1+\int_{r_k}^{r_0} \frac{dr}{r}) = \\
\frac{n_0}{n_k} \sin{\alpha_0} (1+\ln{\frac{r_0}{r_k}}).
\label{l10}
\end{eqnarray}
Presuming now $r_0/r_k \approx 1$, we get:
\begin{equation}
\sin{\alpha_k}=\frac{r_0 n_0 \sin{\alpha_0}}{r_k n_k}.
\label{l11}
\end{equation}
Following the geometry from Fig.~4 we get
\begin{equation}
\frac{\Delta r}{r_i} \tan{\alpha_i}=\frac{\Delta r}{r_i} \frac{1}{\sqrt{\frac{1}
{\sin^2{\alpha_i}}-1}},
\label{l12}
\end{equation}
which in the $\Delta r \rightarrow 0$ and $\alpha_i=\alpha(r)$ continuous limit yields:
\begin{equation}
d\theta=\frac{dr}{r}\frac{1}{\sqrt{\frac{1}{\sin^2{\alpha(r)}}-1}}.
\label{l13}
\end{equation}
Denoting by $\alpha_M$ the final incident angle at the observer and using (\ref{l11})
we get:
\begin{equation}
d\theta=\frac{dr}{r} \frac{1}{\sqrt{\left [\frac{n(r)}{n(r_M)} \right ]^2 \frac{1}{\sin^2{\alpha_M}}
\frac{r}{r_M} -1 }}.
\label{l14}
\end{equation}
By using the fact that $\sin^2{\alpha_M}=\cos^2{\delta}$ we can finally determine
the angle $\theta_r$ as a function of $\delta$ by integrating (\ref{l14}):
\begin{equation}
\theta_r=\int_{z_0+R_E}^{z'+R_E} \frac{dz}{\sqrt{\left [\frac{n(z)}{n(z_0)}\right ]^2 \frac{1}{\cos^2{\delta}}
\frac{z+R_E}{z_0+r_E} -1}} .
\label{l15}
\end{equation}
The above integral can be only numerically computed. There is of course a singularity at $z=z_0$,
which can be eliminated by a Gauss-Chebisev expansion, or by adjusting the step in
the numerical integration in the vicinity of the singularity (the method followed by us).
Computing numerically $\theta_r$ as a function of $\delta$ and by getting from (\ref{approx1})
$\theta$ as a function of $\delta$, we are able now to calculate from (\ref{approx2}) the desired
$\delta_{dev}(\delta)$ dependence.

It is important to mention here that applying the method for $\delta<0$ is
not straightforward.
One must first find in this case the closest point $C$ of the light-ray trajectory relative
to the Earth's surface. This can be done by using the fact that the trajectory is symmetric
in the vicinity of this point. Then, we decompose the trajectory in two parts, the first part
is from $M$ to $C$, and the second from $C$ to $S_r$. The light-ray can be followed now
by the presented methods both on the $CS_r$ and $CM$ segments. The deviation angle can be
also computed.

\subsection{Using the Fermat principle}

The Fermat principle states that light travels between two point along that path
which requires the least time, as compared to other nearby paths. We can of course
reformulate the  Fermat principle by using the optical path instead of time. If the
trajectory of a light-ray travelling in the $X-O-Y$ plane is described by the $y=y(x)$
curve (see the geometry in Fig.~5), we have that:
\begin{eqnarray}
\nonumber s=\int_{P_{ini}(x_0,y_0)}^{P_{fin}(x_f,y_f)} n[x,y(x)] ds = \\
\int_{x_0}^{x_f} n[x,y(x)] \sqrt{1+y'(x)^2} dx,
\label{f1}
\end{eqnarray}
should have a local minima. In the above formula $n[x,y(x)]$ denotes the refractive index of the
medium in point with coordinates $x,y(x)$. In our case $n$ has a spherical symmetry, depending on
the $z=\sqrt{x^2+y(x)^2}-R_E$ altitude only.

\begin{figure}[htb]
\epsfig{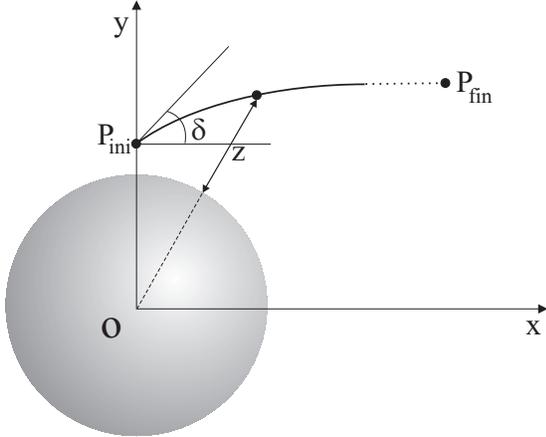}
\caption{Geometry and notations for the method based on the Fermat principle.}
\label{fig5}
\end{figure}

One can write then (\ref{f1}) as:
\begin{equation}
s=\int_{x_0}^{x_f} f[x,y(x),y'(x)] dx ,
\label{f2}
\end{equation}
with:
\begin{equation}
f[x, y(x), y'(x)]=n[\sqrt{x^2+y(x)^2}-R_E] \sqrt{1+y'(x)^2}.
\label{f3}
\end{equation}
We are looking now for the $y(x)$ function that minimizes $s$. If we fix the two
points $P_{ini}$ and $P_{fin}$ between which the light travels, the minima of $s$ leads to a classical
variational problem
\begin{equation}
\delta s=0,
\label{f4}
\end{equation}
with:
\begin{eqnarray}
\delta[y(x)]\mid_{x=x_0} = 0,
\label{f5}  \\
\delta[y(x)] \mid_{x=x_f} =0
\label{f6}.
\end{eqnarray}
The solution of this problem is well-known \cite{arfken}, and given by the second-order
differential equation:
\begin{equation}
\frac{\partial f}{\partial y}-\frac{d}{dx} \left[ \frac{\partial f}{\partial y'} \right] =0.
\label{f7}
\end{equation}
It is straightforward to show that:
\begin{eqnarray}
\frac{\partial f}{\partial y}=\frac{\partial n}{\partial z} \left[ x,y(x) \right]
\frac{y(x)}{\sqrt{y^2(x)+x^2}} \sqrt{1+y'(x)^2},
\label{f8} \\
\nonumber \frac{d}{dx}\left[\frac{\partial f}{\partial y'} \right]=\frac{\partial n}
{\partial z} \left[ x, y(x) \right] \frac{y'(x) y(x)+x}{\sqrt{y(x)^2+x^2}} \times\\
\frac{y'(x)}{\sqrt{1+y'(x)^2}}+
n\left[ x, y(x) \right] y''(x) \frac{1}{[1+y'(x)]^{3/2}}
\label{f9}
\end{eqnarray}
By simple algebra we obtain from this a second-order differential equation
for the $y(x)$ equation describing the trajectory of the light-ray:
\begin{eqnarray}
\nonumber y''(x)=\frac{\partial n}{\partial z} \left[x, y(x) \right]
\frac{[1+y'(x)^2]}{n[x,y(x)] \sqrt{x^2+y(x)^2} }\times \\
\{y(x)[1+y'(x)^2]- y'(x)[y(x)y'(x)+x] \}.
\label{f10}
\end{eqnarray}
Taking the $n(z)$ refractive index profile from our optical
atmosphere model, equation (\ref{f10}) can be numerically
integrated. We start from a $P_{ini}(x_0,y_0)$ point and consider
a $\tan{\delta}=y'(x_0)$ initial derivative. The angle $\delta$
will be the apparent inclination angle of the light-ray in
$P_{ini}$ (the point where the observer is presumed). The $y(x)$
trajectory can be computed by numerically integrating (\ref{f10})
with the $y(x_0)=y_0$ and $y'(x_0)=\tan{\delta}$ initial
conditions. We construct thus the $y(x)$ trajectory from
point-to-point, up to an altitude $z\ge z_s$, where $n(z)=1$ can
be presumed. The derivative of $y(x)$ at this point will determine
the $\delta_r$ angle of the light-ray at the border of the optical
atmosphere.  For altitudes higher than this, the trajectory of the
light-ray is presumed recti-linear. The deviation angle is then
simply approximated as $\delta_{dev}=\delta-\delta_r$, and the
desired $\delta(\delta_{dev})$ dependence can be numerically
computed.

\subsection{The simulation method}

As its name suggests this is simply a computer simulation method
in which we follow the light-ray by segments of infinitesimally
small and fixed-length lines. The angle between two elementary
line is given by estimating the refractive index at their
end-points and by using Snell's law. The trajectory of the
light-ray is then constructed starting from the observer with an
initial angle $\delta$, and computing the ray-path from
point-to-point until it leaves the optical atmosphere and the
refractive index can be taken as $1$. At this point the direction
of the light-ray determines the final angle $\delta_r$, and the
deviation is computed as $\delta_{dev}=\delta-\delta_r$.
Resembling the method based on the Fermat principle, this
simulation also works smoothly for arbitrary positive or negative
$\delta$ values. Due to the fact that for an elementary step the
changes in the refractive index is rather small, it is crucial to
work with the maximal precision offered by the computing
environment.

We mention here that all three methods give identical results, proving their applicability and
our theoretical considerations. As an example, for the parameters of the U.S. Standard
Atmosphere, results for the $\delta_{dev}$ deviation angle versus the apparent
inclination angle, $\delta$, is plotted on Fig.6 ($T_0=10^0 C$, $P_0=1atm=101.3kPa$, $\nu=20000cm^{-1}$
wave-number for the light and $\lambda=6.5K/km$ lapse-rate).

\begin{figure}[htb]
\epsfig{figure=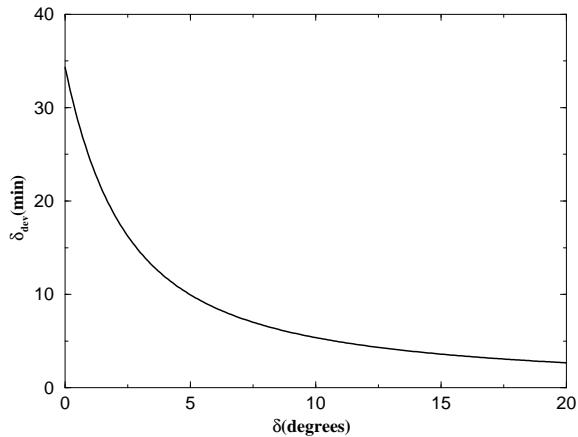,width=3.0in,angle=-0}
\caption{Deviation angle $\delta_{dev}$ between the apparent and real location of a
point-like source far from Earth as a function of the $\delta$ apparent inclination angle.
($T_0=10^0C$, $P_0=1atm$, standard $\lambda=6.5K/km$ lapse-rate  and observation from sea-level)}
\label{fig6}
\end{figure}

These results are
in excellent agreement with the one given by Thomas and Joseph \cite{hopkins} and
the report of the U.S. Naval Observatory (1993) \cite{naval}. From Fig.~6 we learn
that the deviation angle is usually quite small, and becomes important only when viewing
objects in the vicinity of the optical horizon. The deviation angle increases sharply
for very small inclination angles. This effect is responsible for the flattened shape of the
rising (or setting) Sun and Moon, and also for the fact that these heavenly objects appear
more flattened at their bottom. It is also interesting to note here
that the angular extent of the Sun and full-Moon is 32 arcmin, and the maximum
deviation obtained for standard conditions is 34.5 arcmin. This leads us to the observation
that the Sun or Moon is visible even when in reality it is below the geometrical horizon.

\section{Computing the flatness}

Once we numerically determined the $\delta_{dev}$ deviation angle
as a function of the $\delta$ apparent inclination angle, it is
easy to characterize the apparently flat rim of the setting Sun.
Atmospheric refraction influences only the $\Delta_v$ vertical
angular extent of the Sun (or Moon), which becomes thus smaller
than the $\Delta_h$ horizontal extent. The asymmetry ratio (or the
flatness) for the rim of the Sun can be described by the
\begin{equation}
\alpha=\frac{\Delta_h}{\Delta_v},
\label{flatness}
\end{equation}
ratio (of course $\alpha\in [1,\infty)$).
Since both the Sun and the full-Moon are normally
visible under a $\Delta_0=32\: arcmin$ angular extent, we have
$\Delta_h=\Delta_0$. The
value of $\Delta_v$ can be derived after numerically computing
the $\delta$ apparent inclination angle as a
function of the $\delta_r=\delta+\delta_{dev}$ real inclination, i.e. $\delta=F(\delta_r)$.
If the apparent inclination angle for the bottom
of the Sun is $\delta_0$, corresponding to a $\delta_{r0}$ real inclination angle, than
\begin{equation}
\Delta_v(\delta_0)=F(\delta_{r0}+\Delta_0)-\delta_0,
\label{vertical}
\end{equation}
and we get:
\begin{equation}
\alpha(\delta_0)=\frac{\Delta_0}{F(\delta_{r0}+\Delta_0)-\delta_0}
\label{alpha}
\end{equation}
For fixed observation altitude and meteorological conditions
the $\alpha_c$ maximal possible flatness corresponds to the
situation when the bottom of the Sun
touches the horizon. This happens for a $\delta_0=\delta_c$
critical inclination angle.

By decreasing the $\delta$ angle in small steps, and applying the
previously described methods for computing the light-ray
trajectory, both the $\delta=F(\delta_r)$ function and $\delta_c$
is numerically determined. In our calculations we have chosen to
decrease $\delta$ in steps of $0.01^0$.

\section{Results}

Applying the methods presented in the previous sections and our
optical atmosphere model, we systematically computed the
$\alpha_c$ asymmetry ratio for the rim of the setting Sun as a
function of observational altitude and meteorological conditions
parameterized by pressure and temperature. We have also shown that
the value of $\alpha$ is insensitive to details of the considered
optical atmosphere model, proving the stability of our results. In
the following we detail our findings. For all the calculations we
considered monochromatic light with wave-number of
$\nu=20000cm^{-1}$, corresponding to the green color. If otherwise
not specified we considered the height of the troposphere
$z_t=14km$ and the height of the stratosphere as $z_s=50km$.

\subsection{Stability regarding the considered optical atmosphere model}

As discussed in section II., for the construction of the
refractive index profile we borrowed results from the U.S.
Standard Atmosphere model. The main parameters needed by us is the
$z_t$ height of the troposphere, the $z_s$ height of the
stratosphere and the $\lambda$ lapse-rate.

\begin{figure}[htb]
\epsfig{figure=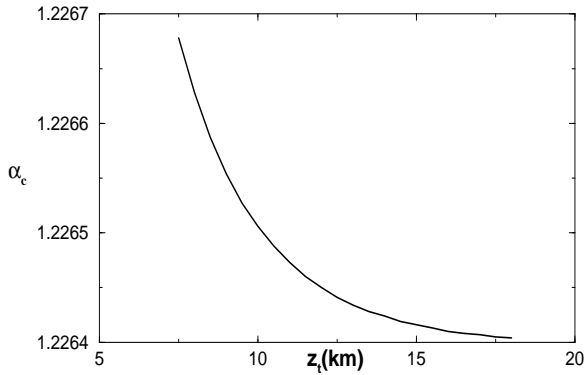,width=3.0in,height=2.0in,angle=-0}
\caption{Maximal observable flatness, $\alpha_c$ as a function of the $z_t$
height of the troposphere.
($T_0=0^0C$, $P_0=1atm$, normal lapse-rate and observation from sea-level)}
\label{fig7}
\end{figure}

\begin{figure}[htb]
\epsfig{figure=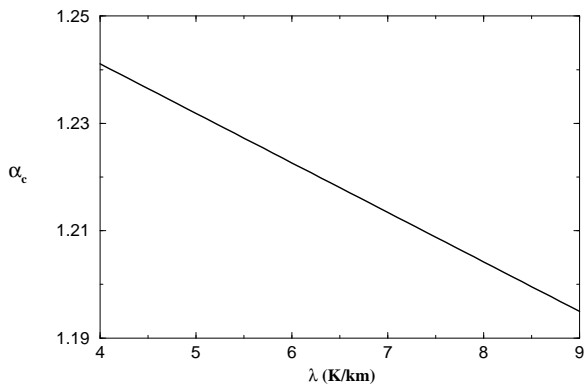,width=3.0in,angle=-0}
\caption{Maximal observable flatness, $\alpha_c$ as a function of the $\lambda$
lapse-rate in the troposphere.
($T_0=0^0C$, $P_0=1atm$, and observation from sea-level)}
\label{fig8}
\end{figure}

It is evident, that the exact value of $z_s$ does not much
influence our results, since in the stratosphere the refractive
index is already very close to $1$. As a first step we studied
thus the influence of $z_t$ on the $\alpha_c$ asymmetry ratio. We
considered normal conditions with $T_0=0^0C$, $P_0=1atm$,
observations at sea-level ($z_0=0$) and $\lambda=6.5K/km$ standard
lapse-rate. As illustrated on Fig.~7, the value of $z_t$ (in a
reasonable range) has no significant influence (note the scale on
the vertical axis). The value of the lapse-rate has already a more
noticeable effect on $\alpha_c$ (Fig.~8), however this variation
is also quite small for the practically important fluctuations
around the standard $\lambda=6.5K/km$ value. We conclude thus,
that our results are quite stable regarding the details of the
considered optical atmosphere model.

\subsection{Asymmetry as a function of the inclination angle}

We computed the observable $\alpha$ flatness as function of the
apparent inclination angle of the Sun's bottom. Results for
$T_0=0^0C$, $P_0=1atm=101.3kPa$, $\lambda=6.5K/km$ and
observations at sea-level are presented on Fig.~9. As emphasized
before, the asymmetric rim of the setting (or rising) Sun becomes
evident only for very small $\delta_0$ values, when the Sun is
close to the horizon. For these normal parameters we get that
$\alpha_c$ is around $1.2$.

\begin{figure}[htb]
\epsfig{figure=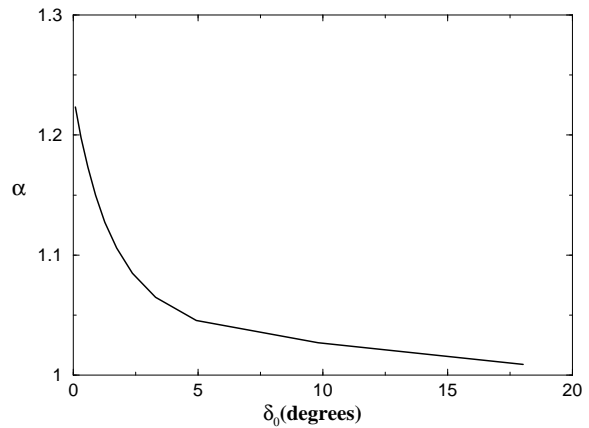,width=3.0in,angle=-0}
\caption{Observable flatness, $\alpha$, as a function of the $\delta_0$
inclination angle of the bottom of the Sun.
($T_0=0^0C$, $P_0=1atm$, observation from sea-level and standard lapse-rate)}
\label{fig9}
\end{figure}

\subsection{Flatness as a function of observation height}

Let us presume now that the observer is at height $z_0$ above the
sea-level, and there is no obstacle in the direction of the
horizon, which is at sea-level (i.e. we are above a vast ocean).
It is obvious that from higher altitude the $\delta_c$ critical
angle will be smaller (becomes negative) and the deviation angle
increases more sharply in the neighborhood of $\delta_c$. This
leads us immediately to the conjecture that the observed flatness
should be larger. Taking the $T_0=0^0C$ and $P_0=1atm$ normal
atmospheric conditions at sea-level, the standard
$\lambda=6.5K/km$, lapse-rate we can compute the $\alpha_c$
maximal observable flatness as a function of observation height.
Results in this sense are plotted in Fig.~10. As expected,
$\alpha_c$ increases with $z_0$. For these normal conditions we
get that from the top of a $5km$ height mountain we would observe
an $\alpha_c\approx 1.5$ flatness and from a commercial flight at
$10km$ height at sunset we would detect an $\alpha_c \approx 1.7$
maximal flatness. For altitudes above $30km$, or observations made
from a space-shuttle we can get extreme values for $\alpha_c$ up
to $2.5$.

\begin{figure}[htb]
\epsfig{figure=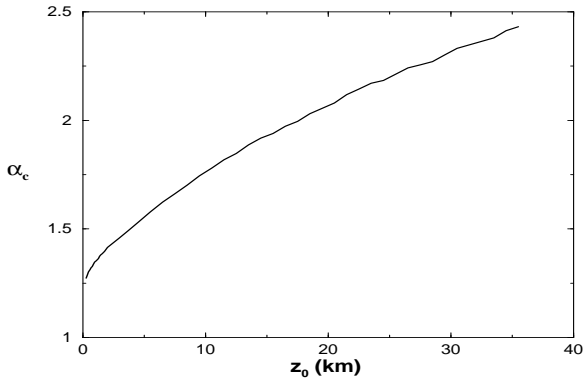,width=3.0in, height=2.0in,angle=-0}
\caption{Maximal observable flatness, $\alpha_c$ as a function of the
$z_0$ observation height.
($T_0=0^0C$, $P_0=1atm$  and standard $\lambda=6.5K/km$ lapse-rate)}
\label{fig10}
\end{figure}

\subsection{Influence of temperature}

The $T_0$ temperature measured at sea-level influences in an
important manner the observed flatness. Computing the temperature
dependence of $\alpha_c$ for observations at sea-level, normal
$P_0=1atm$ atmospheric pressure and standard lapse-rate we get the
values presented in Fig.~11. While for a $T_0=30^0C$ temperature
$\alpha_c$ is around $1.1$ for $T_0=-40^0C$ it becomes $1.3$, and
increases more and more sharply for lower temperatures (arctic
conditions).

\begin{figure}[htb]
\epsfig{figure=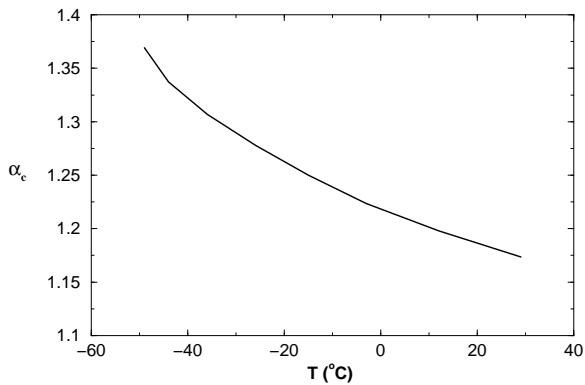,width=3.0in,angle=-0}
\caption{Maximal observable flatness, $\alpha_c$ as a function of the air temperature
measured at sea-level.
($P_0=1atm$, observations at sea-level
and standard $\lambda=6.5K/km$ lapse-rate)}
\label{fig11}
\end{figure}

\subsection{Influence of atmospheric pressure}

Increasing the $P_0$ pressure at sea-level results the increase of
the observable $\alpha_c$ flatness. For the reasonable and
measurable $P_0$ values, $T_0=0^0C$, observation at sea-level and
standard lapse-rate the variation is almost linear (Fig.12). For
an extreme $P_0=125kPa$ pressure one can detect an asymmetry ratio
of $1.3$.

\begin{figure}[htb]
\epsfig{figure=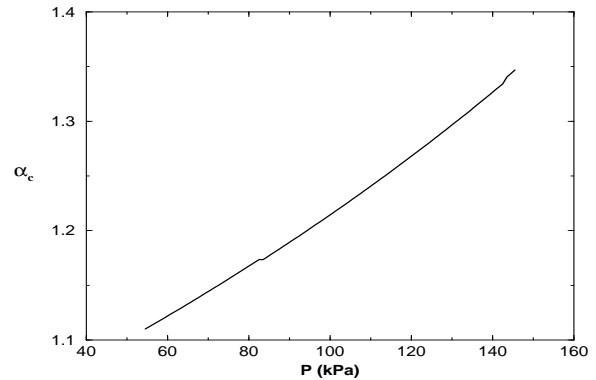,width=3.0in, height=2.0in, angle=-0}
\caption{Maximal observable flatness, $\alpha_c$ as a function of the atmospheric pressure
at sea-level.
($T_0=0^0C$, observation at sea-level
and standard $\lambda=6.5K/km$ lapse-rate)}
\label{fig12}
\end{figure}

\section{Experiments}

By simple experiments it is relatively easy to measure the
flatness of the setting (or rising) Sun. We considered photo and
video experiments, and analyzed the pictures as a function of the
inclination angle of the Sun. With an appropriate filter and
calibrated eye-piece, telescope observations were also possible.
In order to obtain usable pictures with a nicely visible rim we
had to ensure a properly adjusted light intensity. Appropriate
filters, specific atmospheric conditions and usually small
$\delta_0$ inclination angles lead us to usable pictures. When the
Sun is close to the horizon, the inclination angle of the Sun can
be directly determined by analyzing the taken picture. We can use
the fact that the horizontal angular extent of the Sun always
corresponds to $32 arcmin$. The distance of the Sun from the
horizon can then be compared with the horizontal extent of the
setting Sun and the inclination angle results directly from the
picture. Making experiments with the setting or rising Moon is
more complicated, since one needs a full-Moon for this, a work
during the night and we have to deal with pictures where the
horizon is not clearly visible.

First, we have taken several series of pictures in South-Bend
(Indiana, altitude $100m$) both in winter and late-spring,
studying very different temperature conditions. As an immediate
confirmation of our theoretical results, from these pictures it
was obvious that for the same inclination angle the flatness is
bigger in winter, i.e. for lower $T_0$ values. Two series of
digitized and appropriately enhanced pictures are visible on the
web-page dedicated to this study \cite{ourweb}. Results from these
pictures, in comparison with the expected (computed) flatness is
presented on Fig.~13. The first set of pictures is a sunrise in
winter. The mean temperature during the sunrise was $-5^0C$, and
the atmospheric pressure was $103kPa$. It is important to note
that during this sunrise the temperature remained approximately
constant. The second set of picture was made in late-spring for a
sunset. During this sunset the mean-temperature was $12^0C$, the
atmospheric pressure $98kPa$ and the temperature dropped
detectably during the time the pictures were made. Results for the
flatness as a function of the Sun's inclination angle (bottom of
the rim) is plotted with filled circles and triangles for the
winter and spring series, respectively.

\begin{figure}[htb]
\epsfig{figure=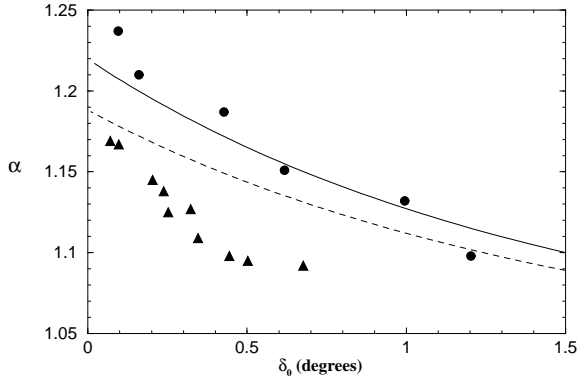,width=3.0in,angle=-0} \caption{Flatness
as a function of inclination angle. Comparison between
experimental results and theory for a sunrise and a sunset photo
sequence. Filled circles are results from sunrise pictures taken
in winter with $T_0=-5^0C$ and $P_0=103kPa$. Filled triangles
correspond for pictures taken in May with $<T_0>=12^0C$ and
$P_0=98kPa$. Observations were made at $z_0=100m$ height with the
horizon roughly at the same altitude. Theoretical results for the
corresponding atmospheric conditions and $\lambda=6.5K/km$
standard lapse-rate is plotted by continuous and dashed lines for
the winter and spring conditions, respectively. } \label{fig13}
\end{figure}

The theoretical curves were constructed for the mentioned mean
temperature, pressure, observation height of $100m$ with the
horizon at the same altitude and a standard lapse-rate. These
results are plotted on Fig.~13 by a continuous and dashed line for
the winter and spring conditions, respectively. As observable from
Fig.~13 for the winter conditions the measured $\alpha$ values are
in acceptable agreement with the one given by our theory. However,
for the spring series the expected $\alpha$ values are higher, and
decreasing slower as a function of the $\delta_0$ inclination
angle, than the measured data. A reason to account for this
sharper trend is the decreasing temperature measured during the
sunset. Taking this effect into account in the calculations, would
definitely result in a trend closer to the observed one. This
temperature variation does not account however, for the constantly
lower values measured for $\alpha$. The only arguments we can give
in this sense is that probably the atmosphere at the time of this
measurement had a refractive index profile different from the one
proposed in our model, behaving in a non-standard manner.

A second set of experiments were realized by video-filming sunsets
in Cluj (Romania). The obtained continuous set of picture allowed
us to follow-up more precisely the flatness as a function of the
inclination angle. The inclination angle was calculated from the
images by the same method as in photographs, i.e. by comparing the
height between the bottom of the Sun and horizon with the
horizontal extent of the setting Sun. Since Cluj is not a flat
region like South-Bend, we had to be careful in choosing the
observation point, and to determine also the altitude of the
optical horizon. Results from video-recording in comparison with
theoretical expectations (corresponding to the appropriate
atmospheric conditions) are plotted in Fig.~14. For this
measurement a quite fair agreement between theoretical and
experimental data is achieved. Since the lapse-rate for the
theoretical prediction was taken from the standard atmosphere
model, the slightly greater $\alpha$ values calculated by us are
explainable.

\begin{figure}[htb]
\epsfig{figure=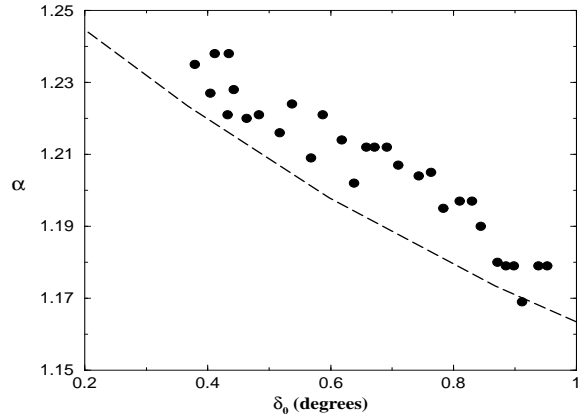,width=3.0in,height=2.2in, angle=-0}
\caption{Experimental (dots) and theoretical results (dashed line) for the flatness as a function
of the inclination angle from a sunset video-recording.
($T_0=7^0 C$ $P_0=108 kPa$, observation at $z_0=400m$, horizon at $z_0=300m$
and a presumed standard $\lambda=6.5K/km$ lapse-rate)}
\label{fig14}
\end{figure}

\section{Pictures and movies from the Internet}

Sunset or sunrise (moon-set or moon-rise) is usually a spectacular
phenomena, and thus it is a favorite theme for professional and
amateur photographers. The Internet is full of beautiful and
useful pictures in this sense. Many of these pictures are taken in
extreme conditions (arctic environment, airplanes, space-shuttle
or high mountains), offering us an excellent possibility to check
our calculations under these conditions, too. Moreover, on the web
there are also interesting, freely downloadable or public domain
movies exemplifying how the observed flatness increases in the
neighborhood of the horizon.

We performed an extensive search on the Internet and collected
non-copyrighted materials about sunset, sunrise, moon-set and
moon-rise. Since their presentation in the context of this paper
is impossible we classified and stored them on the web-page
\cite{ourweb} dedicated to this study. The interested reader can
browse this collection and convince himself or herself that these
pictures support our theoretical results for the estimated
flatness. In agreement with our expectations we found that for a
sunrise (or sunset) viewed from a space-shuttle $\alpha_c$ should
be of the order $2-2.5$, and in arctic environment $\alpha_c$
increases up to values of $1.3$. For most of the everyday, usual
and low-altitude photos one finds $\alpha_c\approx 1.1-1.2$.
Pictures taken from commercial airplane yield $\alpha_c\approx
1.5$. We also found a picture-series for a sunset over the ocean,
where the rim of the Sun is nicely visible. The scenario presented
in this photo is again in agreement with our theoretical
predictions.

The movies stored on our web-page will also convince the reader
about the sharp variation of $\alpha$ as a function of the
$\delta_0$ deviation angle. A public domain movie, showing a
moon-set viewed from the space-shuttle illustrates observations
from high altitudes.

\section{The Green flash}

Although the explanation and the study of the green flash is not the aim
of the present paper, we briefly discuss here how this phenomenon can be understood
and studied through our methods.

From equation (\ref{edlen}) it is obvious that atmospheric
refraction depends on the frequency (color) of the light-ray, a
phenomenon called dispersion. Differently colored light-rays
coming from a source emitting a continuous spectra will suffer
different deviations. More strongly will bend the light-rays with
bigger frequencies, leading to a separation of the colors in the
observed Sun. Since the high-frequency visible components
(corresponding to violet and blue colors) are strongly scattered
by the atmosphere, the green component reaching directly the
observer will suffer the strongest bending. When the Sun
disappears bellow the horizon this component will be observed
lastly, leading to a green-flash on the horizon. Since for
standard atmospheric conditions the dispersion from atmospheric
refraction is tinny, the effect is hardly observable for usual and
everyday conditions. Green flashes become observable exactly under
those conditions under which the flatness is more accentuated.
This means high observation altitudes, low temperature and high
pressure. As an illustration of this, we have just learned that
for airplane pilots it is a quite usual phenomenon. Green flashes
are often seen in non-standard optical atmosphere, when mirages
appear. This is generally the case when layers of air with
strongly different temperatures are in contact.

For a nice presentation and discussion on green flashes the interested reader
should consult the excellent home-page of A.T. Young \cite{young}.

\section{Sunset simulation program}

We also created a computer-code which simulates the sunset on the computer-screen.
By using the discussed simulation method, the program computes and visualizes the rim
of the Sun during a sunset. After fixing the atmospheric conditions (temperature and
pressure) and observation height in the menu, one can follow up how a sunset might look like
in our optical atmosphere. The program
runs under Windows environment and one can freely download the
executable from the web-page accompanying this study \cite{ourweb}.

\section{Conclusions}

Atmospheric refraction is responsible for the asymmetric rim of
the setting (or rising) Sun. Three different methods yielding the
same results were presented here to compute the path of a
light-ray in an optical atmosphere model, where the refraction
index varies continuously as a function of altitude. By
determining the deviation angle between the apparent and real
inclination of a point-like light-source which is at a large
distance from the Earth, we were able to compute the $\alpha$
asymmetry ratio for the rim of the setting Sun. We investigated
$\alpha$ as a function of the inclination angle, observation
altitude and atmospheric conditions. We found that the maximal
flatness obtained  in the vicinity of the horizon increases as a
function of observation altitude and pressure, and decreases as
the temperature increases. We found that $\alpha$ is rather
insensitive to the fine-details of the considered optical
atmosphere model, which makes our results robust. The maximal
flatness observable under normal conditions at sea-level is around
$1.2$. At extremely low temperatures ($-40^0C$) one can observe
values up to $1.3$ and for high altitude observations
(space-shuttle) one can get an asymmetry ratio of $2.5$. By simple
experiments and pictures from the Internet we illustrate and prove
our theoretical predictions. The methods presented here can be
effectively used to study atmospheric refractions in non-standard
atmospheric conditions as well. In this manner one can study
mirages or the green flash phenomenon. A freeware computer program
created by us and downloadable from the web-page accompanying this
study  simulates the sunset for arbitrary atmospheric conditions
and observation altitude.

\section{Acknowledgments}

We thank the Bergen Computational Physics
Laboratory in the framework of the European Community - Access to Research
Infrastructure of the Improving Human Potential programme for access to
their supercomputer facilities.
The work of Z. Neda was sponsored by the Sapientia foundation. We thank I. Albert
and T. N\'eda for useful discussions and help in the video and photo
experiments.


\end{document}